\documentclass[prc,preprint,tightenlines,showpacs]{revtex4}
\usepackage{bm}  
\usepackage{amssymb}
\usepackage{amsmath}
\usepackage{graphicx}  
\begin{document}
%%%%%%%%%%%
%%%%%%%%%%%
\title{Relativistic mean-field study of neutron-rich nuclei}
%%%
\author{B. G. Todd}\email{bonnie@godiva.physics.fsu.edu}
\author{J. Piekarewicz}\email{jorgep@csit.fsu.edu}
\affiliation{Department of Physics, Florida State University, 
	     Tallahassee, FL 32306}
%%%%%%%%%%%
\date{\today} 
\bigskip
\begin{abstract}
 A relativistic mean-field model is used to study the ground-state
 properties of neutron-rich nuclei. Nonlinear isoscalar-isovector 
 terms, unconstrained by present day phenomenology, are added to 
 the model Lagrangian in order to modify the poorly known density 
 dependence of the symmetry energy. These new terms soften the 
 symmetry energy and reshape the theoretical neutron drip line 
 without compromising the agreement with existing ground-state 
 information. A strong correlation between the neutron radius of 
 $^{208}$Pb and the binding energy of valence orbitals is found:
 the smaller the neutron radius of $^{208}$Pb, the weaker the binding 
 energy of the last occupied neutron orbital. Thus, models with the 
 softest symmetry energy are the first ones to drip neutrons. Further, 
 in anticipation of the upcoming one-percent measurement of the neutron 
 radius of $^{208}$Pb at the Thomas Jefferson Laboratory, a close 
 relationship between the neutron radius of $^{208}$Pb and neutron radii 
 of elements of relevance to atomic parity-violating experiments is 
 established.
\end{abstract}
\smallskip
\pacs{21.10.-k,21.10.Dr,21.10.Gv}
\maketitle 

%%%%%%%%%%%%%%%%%%%%%%%%%%%%%%%%%%%%%%%%%%%%%%%%%%%%%%%%%%%
\section{Introduction}
\label{sec:introduction}

Core-collapse supernovae and their remnant neutron stars are an
important source of information pertaining to both superdense matter
and nuclei far from the valley of stability. At present, however,
important open questions remain in the understanding of the structure
and dynamics of neutron stars, such as their size, composition, and
cooling mechanism. Increased knowledge of these issues may in turn
lead to a better understanding of stellar burning and heavy-element
nucleosynthesis.

The structure of spherical neutron stars in hydrostatic equilibrium,
the so-called Schwarszchild stars, is solely determined by the
equation of state of neutron-rich matter in beta equilibrium. Such an
equation of state is well represented as the sum of two distinct
components: {\it i)} a symmetric ($N\!=\!Z$) component that is well
constrained at saturation density plus {\it ii)} a symmetry energy
that accounts for any possible neutron-proton imbalance. Some of the 
important neutron-star properties mentioned above depend critically 
on the poorly known density dependence of the symmetry energy. 
Indeed, as a result of the saturation of symmetric nuclear matter,
most of the pressure supporting the star up to about two times
saturation density is provided by the symmetry energy.

In a set of recent papers, the sensitivity of the symmetry energy to
changes in the model Lagrangian was
investigated~\cite{Ho01a,Ho01b,Ho02,Ca02}. The model Lagrangian was
modified by introducing nonlinear couplings between the isoscalar and
the isovector mesons. These new terms enable one to modify the neutron
skin of heavy nuclei without changing ground-state properties that are
well constrained experimentally. It should be noted that precise
information on the neutron radius of a heavy
nucleus---$^{208}$Pb---might soon become available via a
parity-violating electron scattering experiment at the Jefferson
Laboratory that promises a 1\% accuracy~\cite{Mi99,Ho01c}. While such
a measurement will provide the most accurate determination of the
neutron density of a nucleus to date, it will also impact strongly on
astrophysical observables. Indeed, a model-independent (or
data-to-data) relation between the neutron skin of $^{208}$Pb and
the crust of a neutron star was recently
established~\cite{Ho01a}. This strong correlation emerges as a result
of the similar composition of the neutron skin of a heavy nucleus and
the crust of a neutron star, namely, neutron-rich matter at similar
densities. Further, the measurement of the neutron skin of $^{208}$Pb
constrains the cooling mechanism in neutron stars. A small neutron
radius ($R_n$) suggests a soft symmetry energy which favors a small
proton fraction in dense matter. In turn, a low proton concentration
rules out the enhanced cooling of neutron stars via the direct URCA
process~\cite{Ho02}. If URCA cooling is indeed ruled out, then
observations of enhanced cooling may provide strong evidence in
support of exotic states of matter at the core of neutron stars (see,
for example, Ref.~\cite{Sl02}).

The density dependence of the symmetry energy should also play an
important role on the properties of nuclei far from the valley of
stability. Indeed, some theoretical calculations predict the emergence
of new magic numbers as a result of a reduced spin-orbit splitting
originating from a diffuse neutron skin \cite{Do94,Ot01}. No place is
this more evident than in the neutron-rich oxygen isotopes. The
nucleus of $^{24}$O, which in the traditional scheme has filled
$1d^{5/2}$ and $2s^{1/2}$ orbitals, appears to be the heaviest member
of the isotopic chain that remains stable against particle
emission. Yet, most theoretical calculations predict the existence of
the ``doubly magic'' nucleus $^{28}$O. This issue continues to be
revisited in light of a remarkable experiment that shows that the mere
addition of a single proton stabilizes the fluorine chain up to
$^{31}$F; this is six more neutrons than in $^{24}$O~\cite{Sa99}.

In this---our first---study of neutron-rich nuclei, we focus on the
effect of the new isoscalar-isovector terms on ground-state
properties. This study will be conducted within the framework of
relativistic mean-field models that reproduce a large body of
ground-state observables for nuclei at, or near, the valley of
stability. Yet these observables are mostly insensitive to the density
dependence of the symmetry energy. It is the aim of the present study
to constrain the symmetry energy through a systematic study of
ground-state properties of neutron-rich nuclei. While we recognize
that additional physics, such as pairing, may be needed for more
realistic studies, we limit ourselves to study the impact of the bulk
symmetry energy on the ground-state properties of neutron-rich
nuclei. Further, we focus on self-consistent mean-field models 
in the hope of extending earlier studies of the linear 
response~\cite{Pi00} to incorporate the new isoscalar-isovector
terms.

The manuscript has been organized as follows. In
Sec.~\ref{sec:formalism} we review briefly the relativistic formalism
paying special attention to the role of the new isoscalar-isovector
couplings. We illustrate how these new couplings modify the density
dependence of the symmetry energy while leaving unchanged all
properties of symmetric nuclear matter.  In Sec.~\ref{sec:results} we
present the results of the calculations for various ground-state
observables for a variety of nuclei. A summary and conclusions are
presented in Sec.~\ref{sec:conclusions}.

%%%%%%%%%%%%%%%%%%%%%%%%%%%%%%%%%%%%%%%%%%%%%%%%%
%Put how we solve for stuff\\
%%%%%%%%%%%%%%%%%%%%%%%%%%%%%%%%%%%%%%%%%%%%%%%%%
\section{Formalism}
\label{sec:formalism}

In this section we describe in some detail the model Lagrangian
employed in this work and the mean-field approximation used to compute
ground-state properties for a variety of neutron-rich nuclei. Although
most of the derivations are standard, a review is included here to
illustrate the role of the mixed isoscalar-isovector meson terms first
introduced in Ref.~\cite{Ho01a}. These terms, which supplement the
phenomenologically successful Lagrangian of Ref.~\cite{Mu96}, modify
the density dependence of the symmetry energy, thereby reshaping the
nuclear landscape away from the valley of stability.

\subsection{Effective Lagrangian}
\label{sec:efflagr}
The Lagrangian of Ref.~\cite{Mu96} includes an isodoublet nucleon 
field ($\psi$) interacting via the exchange of one scalar ($\phi$ for 
the sigma) and three vector ($V^{\mu}$ for the omega, ${\bf b}^{\mu}$ 
for the rho, and $A^{\mu}$ for the photon) fields. 
That is,
\begin{eqnarray}
{\mathcal L} &=& 
    \overline{\psi}\left[\gamma^{\mu}
     \left(i\partial_{\mu}-g_{\rm v}V_{\mu}-
     \frac{g_{\rho}}{2}{\bm\tau}\cdot{\bf b}_{\mu}-
     \frac{e}{2}(1+\tau_{3})A_{\mu}\right)-
     \left(M-g_{\rm s}\phi\right)\right]\psi 
     \nonumber \\
     &+&
     \frac{1}{2}\partial^{\mu}\phi\,\partial_{\mu}\phi
      -\frac{1}{2}m_{\rm s}^{2}\phi^{2}
      -\frac{1}{4}V^{\mu\nu}V_{\mu\nu}
      +\frac{1}{2}m_{\rm v}^{2}V^{\mu}V_{\mu} 
     \nonumber \\
     &-&
       \frac{1}{4}{\bf b}^{\mu\nu}\cdot{\bf b}_{\mu\nu}
      +\frac{1}{2}m_{\rho}^{2}\,{\bf b}^{\mu}\cdot{\bf b}_{\mu} 
      -\frac{1}{4}F^{\mu\nu}F_{\mu\nu}
      -U_{\rm eff}(\phi,V^{\mu},{\bf b}^{\mu}) \;, 
\end{eqnarray}
where the various field tensors have been defined as follows:
\begin{subequations}
\begin{eqnarray}
 V_{\mu\nu}&=&\partial_{\mu}V_{\nu}-
              \partial_{\nu}V_{\mu} \;, \\
 {\bf b}_{\mu\nu}&=&\partial_{\mu}{\bf b}_{\nu}-
                    \partial_{\nu}{\bf b}_{\mu} \;, \\
 F_{\mu\nu}&=&\partial_{\mu}A_{\nu}-
              \partial_{\nu}A_{\mu} \;.
\end{eqnarray}
\end{subequations}
In addition to meson-mediated interactions, the Lagrangian is
supplemented with nonlinear meson interactions that serve to simulate
the complicated dynamics that lie beyond the realm of the low-energy
effective theory. Indeed, by fitting the constants that parametrize
these meson terms to the bulk properties of nuclei, rather than to
two-nucleon data, the complicated dynamics originating from nucleon
exchange, short-range effects, and many-body correlations get
implicitly encoded in a small number of empirical constants. For the
purpose of the present discussion, the following local meson terms are
sufficient:
\begin{equation}
    U_{\rm eff}(\phi,V^{\mu},{\bf b}^{\mu})=
	\frac{\kappa}{3!}\Phi^{3}
       +\frac{\lambda}{4!}\Phi^{4}
       -\frac{\zeta}{4!}(W_{\mu}W^\mu)^2 
       -\left(\Lambda_{\rm s}\Phi^{2}
             +\Lambda_{\rm v}W_{\mu}W^\mu
        \right)
        \left({\bf B}_{\mu}\cdot{\bf B}^\mu\right)\;,
 \label{Ueffective}
\end{equation}
where the following definitions have been introduced: 
$\Phi\!=\!g_{\rm s}\phi$,
$W_{\mu}\!=\!g_{\rm v}V_{\mu}$, and 
${\bf B}_{\mu}\!=\!g_{\rho}{\bf b}_{\mu}$.
The inclusion of scalar-meson interactions ($\kappa$ and $\lambda$) 
is dictated by the empirical value of the compression modulus of
symmetric nuclear matter at saturation density
($K\!=\!200\!-\!300$~MeV). In contrast, quartic vector
self-interactions ($\zeta$) affect primarily the high-density
component of the equation of state; their impact at densities 
below normal nuclear-matter saturation density is yet to be
determined~\cite{Mu96}. Finally, the nonlinear mixed 
isoscalar-isovector couplings ($\Lambda_{\rm s}$ and 
$\Lambda_{\rm v}$) are powerful terms because they can be used to 
modify the density dependence of the symmetry 
energy~\cite{Ho01a} while not influencing ground state properties. 
While power counting suggests that other local 
meson terms (such as mixed scalar-vector cubic terms and quartic
rho-meson self-interactions) may be equally important~\cite{Mu96}, 
their phenomenological impact has been documented to be 
small~\cite{Mu96,Ho01a}, so they will be not be considered in this study.

\subsection{Mean-field Equations}
\label{sec:MFE}

The field equations resulting from the above Lagrangian may be solved
exactly in the mean-field limit by replacing all meson-field operators
by their expectation values, which are classical fields~\cite{Se86}.  
For a static, spherically symmetric system this implies 
(using $|{\bf x}|\!\equiv\!r$):
\begin{subequations}
\begin{eqnarray}
 \phi(x) &\rightarrow& \langle\phi(x)\rangle = \phi_{0}(r)\;, \\
  V^{\mu}(x) &\rightarrow& \langle V^{\mu}(x)\rangle =
                                   g^{\mu 0}V_{0}(r) \;,  \\
  {b}^{\mu}_{a}(x) &\rightarrow& \langle b^{\mu}_{a}(x)
                 \rangle = g^{\mu 0}\delta_{a3}\,b_{0}(r)\;, \\  
  A^{\mu}(x) &\rightarrow& \langle A^{\mu}(x)\rangle =
                                   g^{\mu 0}A_{0}(r) \;.  
\end{eqnarray}
\end{subequations}
Similarly, the various baryon sources to which these mesons couple 
must be replaced by their (normal-ordered) expectation values 
in the mean-field ground state. That is,
\begin{subequations}
\begin{eqnarray}
 \overline{\psi}(x){\bf 1}\psi(x) &\rightarrow& 
 \langle
  :\!\overline{\psi}(x){\bf 1}\psi(x)\!:
 \rangle=\rho_{\rm s}(r)\;, \\
 \overline{\psi}(x)\gamma^{\mu}\psi(x) &\rightarrow& 
 \langle
  :\!\overline{\psi}(x)\gamma^{\mu}\psi(x)\!:
 \rangle=g^{\mu 0}\rho_{\rm v}(r)\;, \\
 \overline{\psi}(x)\gamma^{\mu}\tau_{a}\psi(x) &\rightarrow& 
 \langle
  :\!\overline{\psi}(x)\gamma^{\mu}\tau_{a}\psi(x)\!:
 \rangle=g^{\mu 0}\delta_{a3}\rho_{3}(r)\;, \\
 \overline{\psi}(x)\gamma^{\mu}\tau_{p}\psi(x) &\rightarrow& 
 \langle
  :\!\overline{\psi}(x)\gamma^{\mu}\tau_{p}\psi(x)\!:
 \rangle=g^{\mu 0}\rho_{p}(r)\;.
\end{eqnarray}
\end{subequations}
Note that the proton isospin projection operator has been 
defined as $\tau_{p}\!=\!(1\!+\!\tau_{3})/2$. The baryon
sources generate (classical) meson fields that satisfy 
coupled, nonlinear Klein-Gordon equations of the following 
form:
\begin{subequations}
\begin{eqnarray}
 &&
 \left(
  \frac{d^{2}}{dr^{2}}+\frac{2}{r}\frac{d}{dr}-m_{\rm s}^{2}
 \right)\Phi_{0}(r) - g_{\rm s}^{2}
 \left(
  \frac{\kappa}{2}\Phi_{0}^{2}(r)  +
  \frac{\lambda}{6}\Phi_{0}^{3}(r) -
  2\Lambda_{\rm s}B_{0}^{2}(r)\Phi_{0}(r)
 \right)=-g_{\rm s}^{2}\rho_{\rm s}(r)\;, \phantom{xxx} \\
 &&
 \left(
  \frac{d^{2}}{dr^{2}}+\frac{2}{r}\frac{d}{dr}-m_{\rm v}^{2}
 \right)W_{0}(r) -  g_{\rm v}^{2}
 \left(
  \frac{\zeta}{6}W_{0}^{3}(r) + 
  2\Lambda_{\rm v}B_{0}^{2}(r)W_{0}(r)
 \right)=-g_{\rm v}^{2}\rho_{\rm v}(r)\;, \\
 &&
 \left(
  \frac{d^{2}}{dr^{2}}+\frac{2}{r}\frac{d}{dr}-m_{\rho}^{2}
 \right)B_{0}(r) - 2 g_{\rho}^{2}
 \left(
  \Lambda_{\rm s}\Phi^{2}(r)+\Lambda_{\rm v}W_{0}^{2}(r)
 \right)B_{0}(r)
 =-\frac{g_{\rho}^{2}}{2}\rho_{\rm 3}(r)\;. 
\end{eqnarray}
\label{KleinGordonEqn}
\end{subequations}
The photon field couples only to the (point) proton density and 
its solution is thus reduced to quadratures
\begin{equation}
  A_{0}(r)=e\left[
   \frac{1}{r}\int_{0}^{r} dx\,x^{2}\rho_{\rm p}(x)+
   \int_{r}^{\infty} dx\,x\,\rho_{\rm p}(x)\right]\;.
\end{equation}

The eigenstates of the Dirac equation, for the spherically symmetric 
mean-field ground state assumed here, may be classified according to 
a generalized angular momentum $\varkappa$. Thus, the single-particle
solutions of the Dirac equation may be written as 
\begin{equation}
 {\cal U}_{n \varkappa mt}({\bf x}) = \frac {1}{r}
 \left( \begin{array}{c}
   \phantom{i}
   g_{n \varkappa t}(r) {\cal Y}_{+\varkappa \mbox{} m}(\hat{\bf x})  \\
  if_{n \varkappa t}(r) {\cal Y}_{-\varkappa \mbox{} m}(\hat{\bf x})
 \end{array} \right)\zeta_{t},
\label{Uspinor}
\end{equation}
where $\zeta_{t}$ denotes a two-component (Pauli) spinor in isospin
space (with $t\!=\!\pm\!1/2$ for protons and neutrons, respectively),
$n$ and $m$ are the principal and magnetic quantum numbers,
respectively, and the spin-spherical harmonics are defined as
\begin{equation}
 {\cal Y}_{\varkappa\mbox{} m}(\hat{\bf x}) \equiv
 \langle{\hat{\bf x}}|l{\scriptstyle\frac{1}{2}}jm\rangle\;; \quad
 j = |\varkappa| - \frac {1}{2} \;; \quad
 l = 
     \begin{cases}
           \varkappa\;,  & {\rm if} \; \varkappa>0; \\
        -1-\varkappa\;,  & {\rm if} \; \varkappa<0. 
     \end{cases}
 \label{curlyy}
\end{equation}
Note that the phase convention adopted in Eq.~(\ref{Uspinor}) 
({\it i.e.,} the relative factor of $i$) is such that real 
bound-state wave functions ($g$ and $f$) are generated if 
the mean-field potentials are also real. Further, the 
following spinor normalization has been adopted:
\begin{equation}
 \int d^{3}x\,{{\cal U}}^{\dagger}_{\alpha}({\bf x})
                      \,{\cal U}_{\alpha}({\bf x})=
 \int_{0}^{\infty} dr\,\Big(g_{a}^{2}(r)+
			    f_{a}^{2}(r)\Big)=1\;,
\end{equation}
where $\{\alpha\}\!\equiv\!\{a\!=\!n\varkappa t;m\}$ denotes
the collection of all quantum numbers required to describe
the single-particle Dirac spinor. It then follows that the 
coupled differential equations satisfied by the radial 
components of the Dirac spinor are given by
\begin{subequations}
\begin{eqnarray}
  &&
  \left(\frac{d}{dr}+\frac{\varkappa}{r}\right)g_{a}(r)
 -\left[E_{a}+M-\Phi_{0}(r)-W_{0}(r)\mp\frac{1}{2}B_{0}(r)
 -e\left\{
    \begin{array}{c}  1 \\  0 \end{array}
  \right\}A_{0}(r)
  \right]f_{a}(r)=0\;, \phantom{xxxx} \\
  &&
  \left(\frac{d}{dr}-\frac{\varkappa}{r}\right)f_{a}(r)
 +\left[E_{a}-M+\Phi_{0}(r)-W_{0}(r)\mp\frac{1}{2}B_{0}(r)
 -e\left\{
    \begin{array}{c}  1 \\  0 \end{array}
  \right\}A_{0}(r)
  \right]g_{a}(r)=0\;, 
\end{eqnarray}
\label{DiracEqn}
\end{subequations}
where the upper and lower numbers in these equations correspond to
protons and neutrons, respectively. Having determined all occupied
single-particle states, the various ground-state densities, which 
act as sources for the meson fields in the Klein-Gordon equations 
[see Eq.~(\ref{KleinGordonEqn})], may now be computed. They are 
given by
\begin{subequations}
\begin{eqnarray}
  &&
  \rho_{\rm s}(r)=\rho_{\rm s,p}(r)+\rho_{\rm s,n}(r)\;,
  \label{RhoSigma} \\
  &&
  \rho_{\rm v}(r)=\rho_{\rm v,p}(r)+\rho_{\rm v,n}(r)\;,
  \label{RhoOmega} \\
  &&
  \rho_{3}(r)=\rho_{\rm v,p}(r)-\rho_{\rm v,n}(r)\;,
  \label{RhoRho0} \\
  &&
  \rho_{\rm p}(r)=\rho_{\rm v,p}(r)\;,
  \label{RhoGamma} 
\end{eqnarray}
\end{subequations}
where scalar and vector densities have been defined as
\begin{equation}
  \left(
    \begin{array}{c}
      \rho_{\rm s,t}(r) \\
      \rho_{\rm v,t}(r) 
    \end{array}	
   \right) = \sum_{n\varkappa}^{\rm occ}
   \left(\frac{2j_{\varkappa}+1}{4\pi r^{2}}\right)
   \Big(g_{n\varkappa t}^{2}(r)\mp f_{n\varkappa t}^{2}(r)\Big) \;. 
  \label{RhoSV} 
\end{equation}

Ground-state properties of the system are described by a solution 
of the coupled, ordinary differential equations for the classical 
meson fields [Eq.~(\ref{KleinGordonEqn})] and for the Dirac 
single-particle states [Eq.~(\ref{DiracEqn})]. The solution must 
be self-consistent, that is, the meson fields generating the Dirac 
mean-field potentials must satisfy Klein-Gordon equations having 
ground-state densities constructed from the same single-particle 
states as their sources. Thus, an iterative procedure must be
implemented. The self-consistent procedure starts with initial 
Woods-Saxon shaped meson fields of reasonable strength and range 
to generate, via a conventional Runge-Kutta algorithm, bound-state 
energies and corresponding wave functions for all occupied 
single-particle states. At this point, scalar and vector 
densities ($\rho_{\rm s,t}$ and $\rho_{\rm v,t}$) are computed. 
Using these as sources for the meson-field equations, new meson 
fields are generated by using Green's function techniques.
The newly generated meson fields will differ, in general, from
the initial Woods-Saxon guess. Thus, this iterative procedure
must continue until self-consistency (convergence) is achieved. 

\section{Results}
\label{sec:results}

The symmetry energy of infinite nuclear matter impacts on the dynamics
of neutron-rich nuclei. After all, the symmetry energy describes how
the energy of nuclear matter increases as the system departs from
equal numbers of neutrons and protons. To investigate the structure of
neutron-rich nuclei we use a variety of effective field theory (EFT)
models that differ in their prediction for the density dependence of
the symmetry energy.

Three models will be considered in this text: the very successful 
NL3~\cite{La97,La99} along with the newer S271 and Z271~\cite{Ho01a} 
parameter sets. The S271 and Z271 models are alike the NL3 in that they 
are constrained to the following properties of symmetric nuclear 
matter: {\it i)} nuclear saturation at a Fermi momentum of 
$k_{\rm F}\!=\!1.30$~fm$^{-1}$, {\it ii)} a binding energy per 
nucleon of $16.24$ MeV, and {\it iii)} a compression modulus of 
$K\!=\!271$~MeV. The first table (Table~\ref{Table1}) lists the 
various parameter sets that are needed to reproduce these properties 
of symmetric nuclear matter at the mean-field level.

%%%%%%%%%%%%%%%%%%%%%%%%%%%%%%%%%%%%%%%%%%%%%%%%%%%%%%%%%%%%%%%%%
\begin{table}
\caption{Model parameters used in the calculations. The
parameter $\kappa$ and the scalar mass $m_{\rm s}$ are
given in MeV. The nucleon, rho, and omega masses are kept
fixed at $M\!=\!939$, $m_{\rho}\!=\!763$, and
$m_{\omega}\!=\!783$~MeV, respectively---except in the
case of the NL3 model where it is fixed at
$m_{\omega}\!=\!782.5$~MeV.}
\begin{ruledtabular}
\begin{tabular}{lccccccc}
 Model & $m_{\rm s}$  & $g_{\rm s}^2$ & $g_{\rm v}^2$ &
         $g_{\rho}^2$ & $\kappa$ & $\lambda$ & $\zeta$ \\
 \hline
 NL3  & 508.194 & 104.3871  & 165.5854 & 79.6000 
      & 3.8599 & $-$0.01591 &   0.00 \\
 S271 & 505.000 &  81.1071  & 116.7655 & 85.4357 
      & 6.6834 & $-$0.01580 &   0.00 \\
 Z271 & 465.000 &  49.4401  &  70.6689 & 90.2110
      & 6.1696 & $+$0.15634 &   \ 0.06
\label{Table1}
\end{tabular}
\end{ruledtabular}
\end{table}
%%%%%%%%%%%%%%%%%%%%%%%%%%%%%%%%%%%%%%%%%%%%%%%%%%%%%%%%%%%%%%%%%

\begin{figure*}[ht]
\begin{center}
\includegraphics[width=4.0in,angle=0,clip=true]{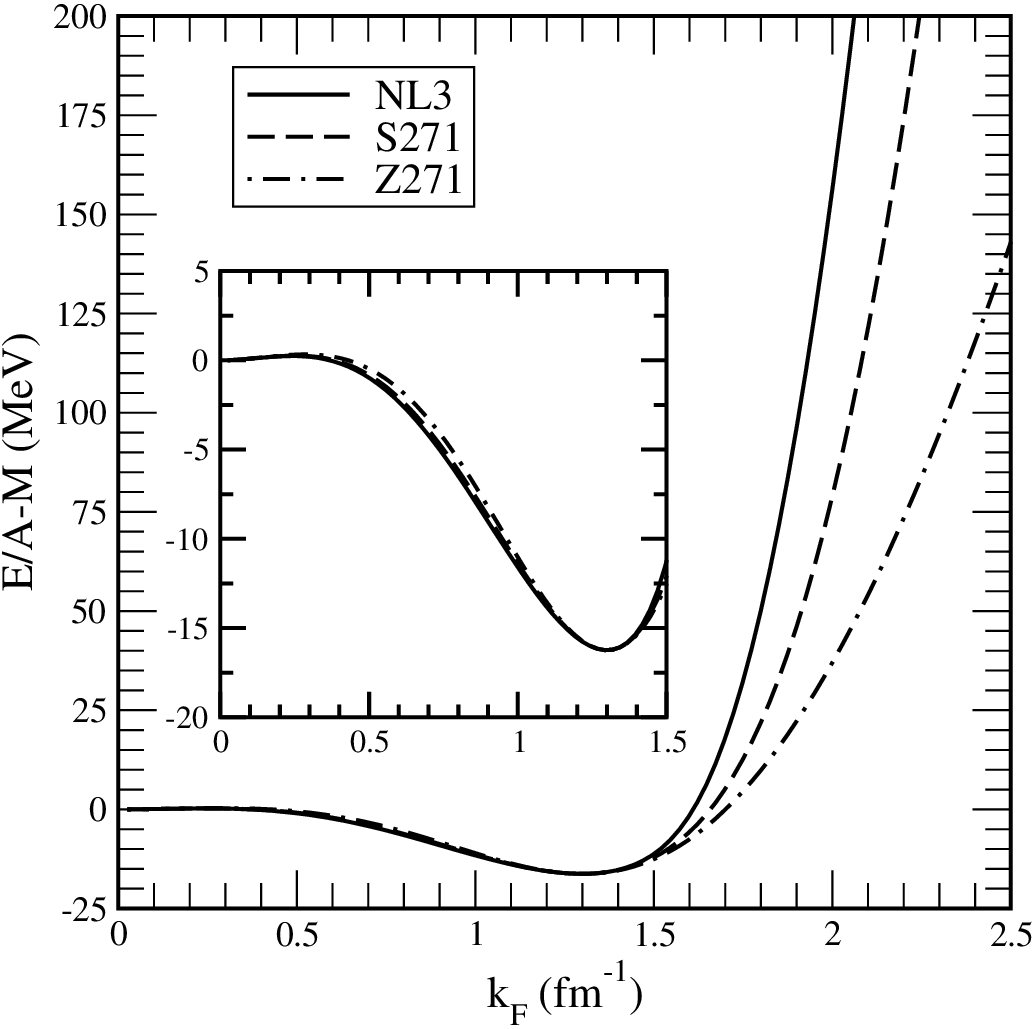}
\caption{Equation of state of symmetric nuclear matter
         for the three models considered in the text.} 
\label{Figure1}
\end{center}
\end{figure*}

In Fig.~\ref{Figure1} the equation of state for symmetric nuclear
matter is displayed for the three models discussed in the text. 
That all models are identical at (and near) saturation density 
follows from the fitting procedure that has produced the above 
mentioned constraints. While significant discrepancies among the 
models emerge at high density ($k_{\rm F}\!\agt\!1.50$~fm$^{-1}$), 
presumably due to differences in the input values for the effective 
nucleon mass $M^{*}$ and $\zeta$, these discrepancies disappear 
at the densities ($k_{\rm F}\!\alt\!1.30$~fm$^{-1}$) relevant to 
the physics of finite nuclei (see inset in the figure). Thus, the 
equation of state for symmetric nuclear matter is model independent 
in this range and the only bulk property of infinite nuclear matter 
that can lead to a model dependence in ground-state observables of 
finite nuclei is the symmetry energy.

%%%%%%%%%%%%%%%%%%%%%%%%%%%%%%%%%%%%%%%%%%%%%%%%%%%%%%%%%%%%%%%%%
%%%%%%%%%%%%%%%%%%%%%%%%%%%%%%%%%%%%%%%%%%%%%%%%%%%%%%%%%%%%%%%%%

Unfortunately, the density dependence of the symmetry energy is 
poorly known. Indeed, even the symmetry energy at saturation 
density is not well constrained experimentally. It is some
average between the symmetry energy at saturation density and the
surface symmetry energy that is constrained by the binding energy 
of nuclei. As a consequence, we adjust the value of the $NN\rho$ 
coupling constant to reproduce a symmetry energy of 
$a_{\rm sym}\!=\!25.67$~MeV at a Fermi momentum of
$k_F\!=\!1.15$~fm$^{-1}$ 
(or $\rho\!=\!0.10$~fm$^{-3}$)~\cite{Ho01b,Ho02}. (For a recent
discussion on the surface symmetry energy see Ref.~\cite{Da03}.)

%%%%%%%%%%%%%%%%%%%%%%%%%%%%%%%%%%%%%%%%%%%%%%%%%%%%%%%%%%%%%%%%%
\begin{table}
\caption{Symmetry energy at saturation density, binding energy per 
nucleon, root-mean-square charge radius (point-proton radius in 
parenthesis) and neutron radius of ${}^{208}$Pb for the various 
models considered in the text (with $\Lambda_{\rm v}$=0).}
\begin{ruledtabular}
\begin{tabular}{lcclc}
 Model & $a_{\rm sym}^{\rm sat}$~(MeV) 
       & $B/A~$(MeV) & $R_{\rm ch}(R_{p})$~(fm) 
       & $R_{n}$~(fm) \\
 \hline
 NL3  & 37.285     & 7.854 & 5.509 (5.460) & 5.740 \\
 S271 & 36.637     & 7.939 & 5.509 (5.460) & 5.714 \\
 Z271 & 36.298     & 7.775 & 5.508 (5.459) & 5.700 \\
 Exp  & $\sim 37$  & 7.885 & 5.504         &  \ unknown 
\label{Table2}
\end{tabular}
\end{ruledtabular}
\end{table}
%%%%%%%%%%%%%%%%%%%%%%%%%%%%%%%%%%%%%%%%%%%%%%%%%%%%%%%%%%%%%%%%%

In Table~\ref{Table2} predictions for the binding energy per
nucleon~\cite{Au95}, root-mean-square charge 
radius~\cite{Vr87,Fr95}, and
neutron radius of $^{208}$Pb are displayed for the three models
considered in the text. The predictions are within one percent of the
experimental values, except for the neutron radius which is poorly
known. The nonlinear coupling ($\Lambda_{\rm v}$) between the
isoscalar and the isovector mesons [see Eq.~(\ref{Ueffective})]
enables one to modify the density dependence of the symmetry energy,
and thus the neutron radius of $^{208}$Pb, while leaving well-known
ground-state properties intact. This suggests that existing
ground-state information, such as charge densities and binding
energies, do not determine the neutron radius uniquely. Thus,
a new measurement, such as the neutron radius in
$^{208}$Pb~\cite{Mi99},  is needed to provide important constraints on the
density dependence of the symmetry energy. It is therefore the aim of
this paper to correlate ground-state properties of neutron-rich nuclei
to the neutron radius of $^{208}$Pb.

%%%%%%%%%%%%%%%%%%%%%%%%%%%%%%%%%%%%%%%%%%%%%%%%%%%%%%%%%%%%%%%%%
\begin{figure}[ht]
\begin{center}
\includegraphics[width=4.0in,angle=0,clip=true]{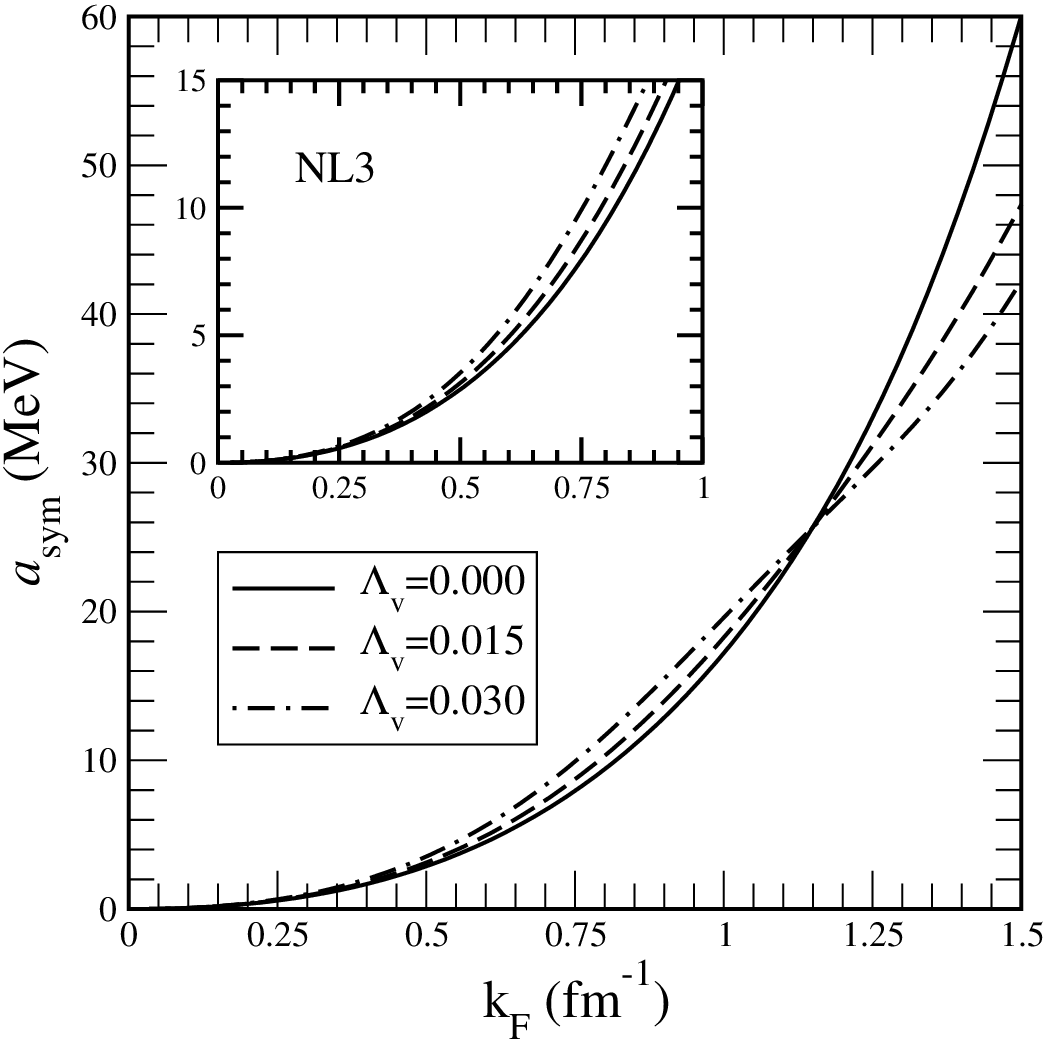}
\caption{The symmetry energy in the NL3 model.}
\label{Figure2}
\end{center}
\end{figure}
%%%%%%%%%%%%%%%%%%%%%%%%%%%%%%%%%%%%%%%%%%%%%%%%%%%%%%%%%%%%%%%%%

The model dependence of the symmetry energy is generated by tuning 
the isoscalar-isovector term. (For simplicity, 
we set the isoscalar-isovector coupling $\Lambda_{\rm s}$ to zero
henceforth.) Note that changing $\Lambda_{\rm v}$ has no effect on
the bulk properties of symmetric nuclear matter as the expectation
value of the rho-meson field is identically zero in the $N\!=\!Z$
limit. It does affect the value of the symmetry energy so a mild 
adjustment of the $NN\rho$ coupling constant is required to keep
the symmetry energy fixed at $a_{\rm sym}\!=\!25.67$~MeV; this 
ensures that the binding energy per nucleon in $^{208}$Pb remains 
fixed at $7.87$~MeV~\cite{Au95}.

In Fig.~\ref{Figure2} the symmetry energy of infinite nuclear 
matter is plotted using the NL3 parameter set for different 
values of the isoscalar-isovector term $\Lambda_{\rm v}$ (the 
original NL3 parametrization has 
$\Lambda_{\rm v}\!\equiv\!0$~\cite{La97}). Increasing 
$\Lambda_{\rm v}$ softens the symmetry energy thereby 
reducing the internal pressure of the system. As a result, the 
original NL3 model predicts both the largest neutron radius in 
${}^{208}$Pb and the largest neutron-star radii (for a given 
mass)~\cite{Ho01a,Ho01b}. The inset in the figure shows the 
symmetry energy at the densities relevant to the dynamics of 
neutron-rich nuclei. Note that because all parameterizations are 
constrained to have the same symmetry energy at
$k_F\!=\!1.15$~fm$^{-1}$, models with a softer symmetry energy 
have a larger symmetry energy at low densities. In these softer 
models there is a higher price to pay for departing from equal 
numbers of protons and neutrons. This suggests a definite 
correlation: models with the smallest neutron skin 
($R_{n}\!-\!R_{p}$) in $^{208}$Pb should be the first ones to 
drip neutrons. 

%%%%%%%%%%%%%%%%%%%%%%%%%%%%%%%%%%%%%%%%%%%%%%%%%%%%%%%%%%%%%%%%%
\begin{figure}[ht]
\begin{center}
\includegraphics[width=4.0in,angle=0,clip=true]{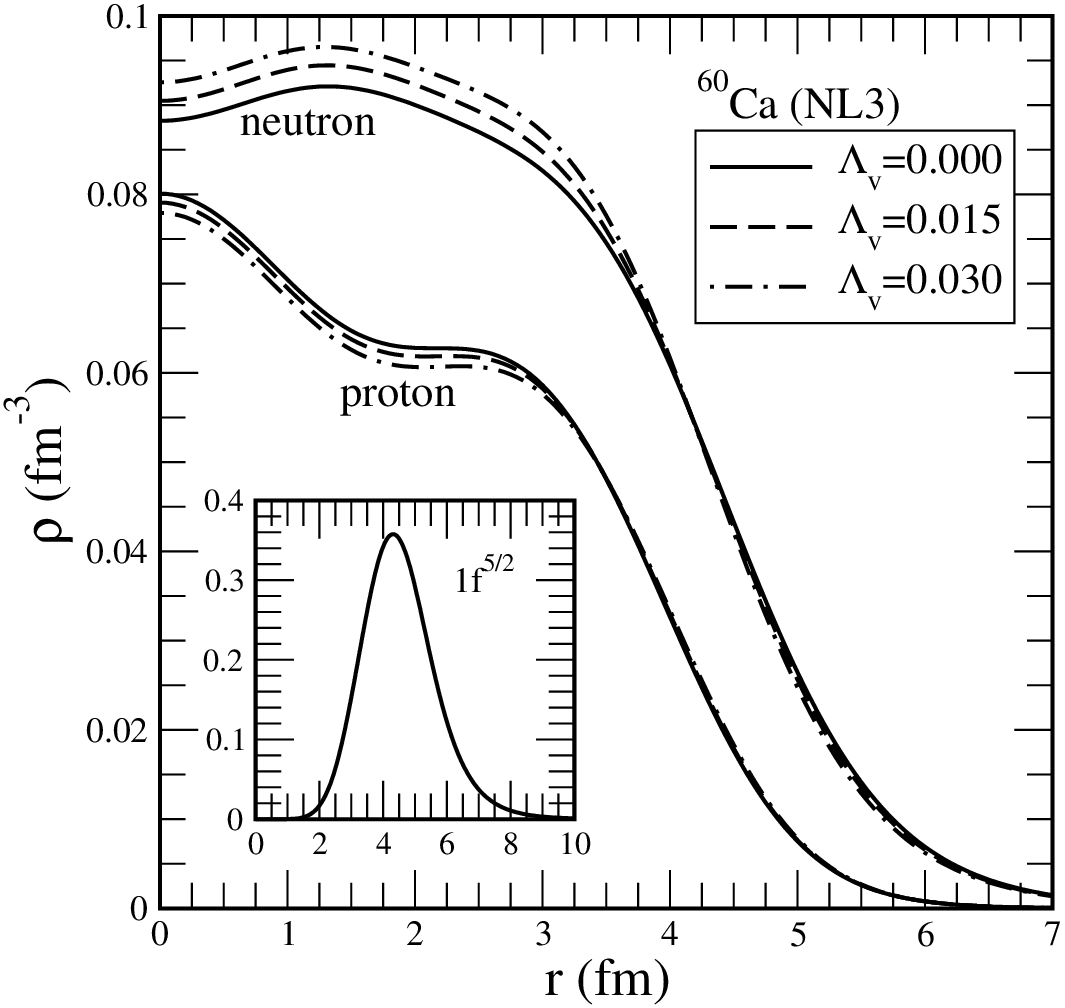}
\caption{Neutron and proton densities for $^{60}$Ca in the 
NL3 model for various values of the isoscalar-isovector term 
$\Lambda_{\rm v}$. The inset shows the surface-peaked nature
of the (square of the) upper component of the least bound 
$1f^{5/2}$-neutron.}
\label{Figure3}
\end{center}
\end{figure}
%%%%%%%%%%%%%%%%%%%%%%%%%%%%%%%%%%%%%%%%%%%%%%%%%%%%%%%%%%%%%%%%%

Further evidence for this behavior is provided in Fig.~\ref{Figure3}
where proton and neutron densities for the neutron-rich nucleus
${}^{60}$Ca are displayed. The nonlinear isoscalar-isovector term
$\Lambda_{\rm v}$ is used to change the density dependence of the
symmetry energy which in turn modifies the neutron radius of
$N\!\ne\!Z$ nuclei. Indeed, the predicted values for the neutron skin
of ${}^{60}$Ca are $R_{n}\!-\!R_{p}\!=\!0.608, 0.567, 0.523$~fm for
$\Lambda_{\rm v}\!=\!0, 0.015, 0.030$, respectively. Note that the
proton radius remains fixed at $R_{p}\!=\!3.562\pm0.013$~fm.  The
inset shows the square of the upper component of the least bound
($1f^{5/2}$) neutron in ${}^{60}$Ca (for $\Lambda_{\rm v}\!=\!0$) to
illustrate how it is most sensitive to the symmetry energy in the
low-density surface region; this is the region (according to the inset
in Fig.~\ref{Figure2}) where one expects the model with the largest
neutron skin to give the strongest binding energy for the $1f^{5/2}$
neutron. In accordance with this statement we obtain a binding energy
for the $1f^{5/2}$ neutron in ${}^{60}$Ca of $4.907, 4.705, 4.443$~MeV
for a corresponding neutron skin of $R_{n}\!-\!R_{p}\!=\!0.608, 0.567,
0.523$~fm, respectively.  

%%%%%%%%%%%%%%%%%%%%%%%%%%%%%%%%%%%%%%%%%%%%%%%%%%%%%%%%%%%%%%%%%
\begin{figure}[ht]
\begin{center}
\includegraphics[width=6in,angle=0,clip=true]{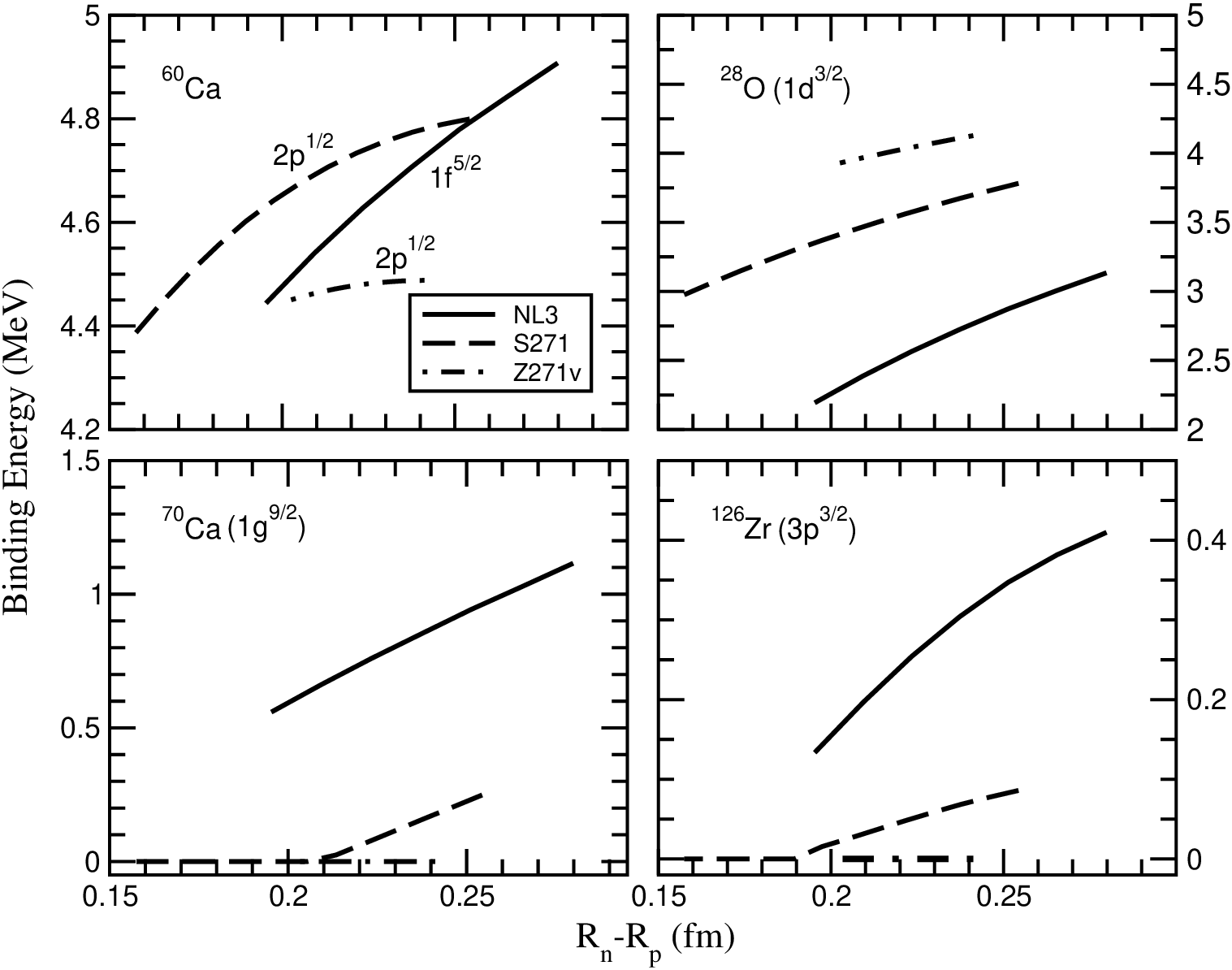}
\caption{The least bound neutron for various neutron-rich
nuclei as a function of the neutron skin in ${}^{208}$Pb
for the three models considered in the text. A ``binding 
energy'' of $0$~MeV indicates that the single-particle
orbital is unbound.}
\label{Figure4}
\end{center}
\end{figure}
%%%%%%%%%%%%%%%%%%%%%%%%%%%%%%%%%%%%%%%%%%%%%%%%%%%%%%%%%%%%%%%%%

This correlation is displayed in graphical form in the upper-left-hand
panel of Fig.~\ref{Figure4}, where the binding energy of the least
bound neutron in $^{60}$Ca is plotted for the NL3 model (solid line)
as a function of the neutron skin in $^{208}$Pb. As evinced by the
other panels in the figure, this correlation holds throughout the
periodic table. Moreover, it is model independent. Indeed, similar
plots have been added for the S271 (dashed line) and Z271 (dot-dashed
line) parameterizations.  In all cases the binding energy of the least
bound neutron increases with increasing neutron skin. Note, however,
that the shell ordering is not constrained among the models. For
$^{60}$Ca the NL3 model predicts the $1f^{5/2}$ orbital to be the
least bound, whereas in the S271 and Z271 models it is the $2p^{1/2}$
orbital which is the least bound. Moreover, this correlation, namely,
that an increase in the isoscalar-isovector coupling is responsible
for a decrease in both the neutron skin of $^{208}$Pb and in the
binding energy of the least bound neutron of a nucleus, should only be
applied within a model and should not be used to compare among
different models. Indeed, the Z271v model predicts the weakest
(strongest) binding energy for the last neutron orbital in ${}^{60}$Ca
(${}^{28}$O) among all the models.

A case of particular interest is the doubly magic nucleus ${}^{28}$O
(upper-right-hand panel of Fig.~\ref{Figure4}) which appears to be 
particle unstable~\cite{Sa99} in spite of the many theoretical
predictions to the contrary~\cite{Wa90,Re95,Kr97,La98}. While the case 
of ${}^{28}$O in particular, and the whole isotopic chain in general, 
deserves special attention and thus a separate publication, suffices 
to say that in all the self-consistent relativistic mean-field models 
considered here, the $1d^{3/2}$ orbital in ${}^{28}$O is predicted to 
be bound by at least 2~MeV. 

We finish the discussion of Fig.~\ref{Figure4} by addressing the
model dependence of the neutron drip lines in calcium and zirconium.
Predictions for the least bound neutron in the neutron-rich nuclei 
${}^{70}$Ca (lower-left-hand panel) and ${}^{126}$Zr (lower-right-hand 
panel) are displayed as a function of the neutron skin in $^{208}$Pb.
Note that a ``binding energy'' of 0~MeV indicates that the neutron is 
unbound and drips. The NL3 model predicts the last occupied neutron 
orbital in both nuclei ($1g^{9/2}$ and $3p^{3/2}$, respectively) to be 
bound, albeit only weakly. This is in contrast to the S271 model for 
which both single-particle orbitals are bound but only for those 
parameterizations with a large neutron skin in $^{208}$Pb; that is, 
with a stiff symmetry energy. The Z271 model predicts both nuclei to 
be particle unstable for all values of $R_{n}\!-\!R_{p}$ in 
$^{208}$Pb.

%%%%%%%%%%%%%%%%%%%%%%%%%%%%%%%%%%%%%%%%%%%%%%%%%%%%%%%%%%%%%%%%%
\begin{figure}[ht]
\begin{center}
\includegraphics[width=4.5in,angle=0,clip=true]{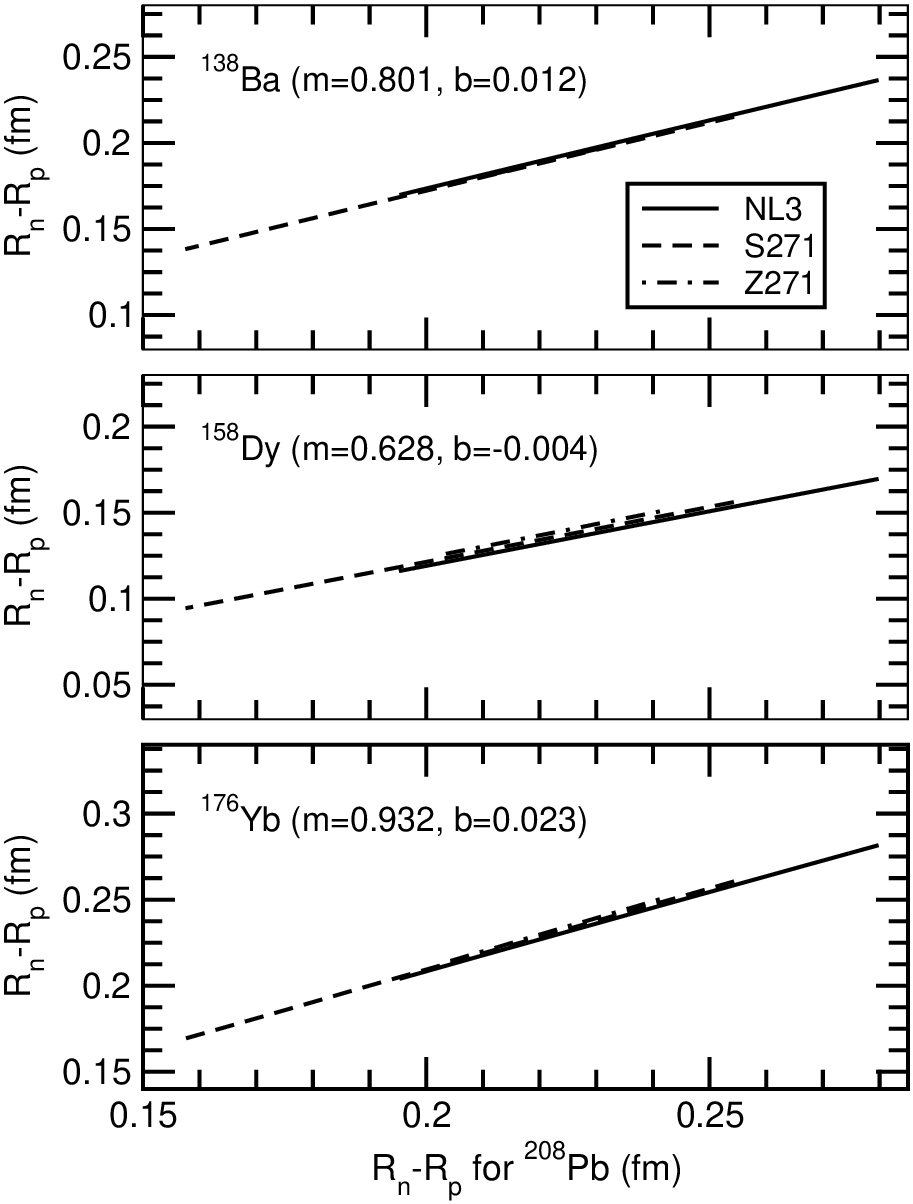}
\caption{Skin-skin correlations for three heavy nuclei of 
possible relevance to atomic parity violation (${}^{138}$Ba,
${}^{158}$Dy, and ${}^{176}$Yb) as a function of the neutron 
skin of $^{208}$Pb for the three models considered in the text. 
Quantities in parenthesis represent linear regression coefficients 
(slope and intercept).}
\label{Figure5}
\end{center}
\end{figure}
%%%%%%%%%%%%%%%%%%%%%%%%%%%%%%%%%%%%%%%%%%%%%%%%%%%%%%%%%%%%%%%%%

We conclude this section with a brief comment on the impact of 
a 1\% measurement of the neutron radius in $^{208}$Pb on the 
neutron radius of other heavy nuclei that have been identified 
as promising candidates for atomic parity-violation experiments:
barium, dysprosium, and ytterbium~\cite{Xi90,De95,
Bu99}. Part of the appeal of these atoms is the existence of very 
close (nearly degenerate) levels of opposite parity that 
considerably enhance parity-violating amplitudes. Unfortunately, 
parity-violating matrix elements are contaminated by uncertainties 
in both atomic and nuclear structure. A fruitful experimental strategy 
for removing the sensitivity to the atomic theory is to measure ratios 
of parity-violation observables along an isotopic chain. This leaves 
nuclear-structure uncertainties, in the form of the neutron radius, 
as the limiting factor in the search for physics beyond the standard 
model~\cite{Po92,Ch93,Mu99}. All three elements, barium, dysprosium, 
and ytterbium, have long chains of naturally occurring isotopes. While 
the experimental strategy demands a precise knowledge of neutron radii 
along the complete isotopic chain, we only correlate here (as means of 
illustration) the neutron radius of $^{208}$Pb to the neutron radius
of a member of the isotopic chain having a closed neutron shell (or 
subshell); we only consider here:
${}^{138}$Ba$(Z\!=\!56; N\!=\!82)$~\cite{Sc02},
${}^{158}$Dy$(Z\!=\!66; N\!=\!92)$, and
${}^{176}$Yb$(Z\!=\!70; N\!=\!106)$. Note that for the open 
proton shell a spherical average is performed. That is, the
factor of $2j_{k}+1$ in Eq.~(\ref{RhoSV}) is simply replaced
by the actual number of protons in the orbital.

The neutron skins of ${}^{138}$Ba, ${}^{158}$Dy, and ${}^{176}$Yb,
are correlated to the corresponding neutron skin of $^{208}$Pb in 
the three panels of Fig.~\ref{Figure5}. We observe a tight linear 
correlation that is largely model independent. The linear regression 
coefficients (slope $m$ and intercept $b$) have been enclosed in 
parenthesis. A theoretical spread of approximately $0.3$~fm in the 
neutron radius of $^{208}$Pb was estimated in Refs.~\cite{Po92,Fu02} 
on the basis of best-fit models. Most of this spread is driven by 
the difference between relativistic and nonrelativistic models, which 
has recently been attributed to the poorly known density dependence 
of the symmetry energy~\cite{Pi02}. With the culmination of the the 
Parity Radius Experiment at the Jefferson Laboratory~\cite{Mi99}, the 
theoretical spread will be replaced by a genuine experimental error 
that is five times smaller, that is, 
$\Delta R_{n}(^{208}{\rm Pb})\simeq 0.056$~fm. This 1\% measurement 
of the neutron radius in $^{208}$Pb translates into a neutron radius 
uncertainty of 
$\Delta R_{n}({}^{138}{\rm Ba})\simeq 0.045$~fm,
$\Delta R_{n}({}^{158}{\rm Dy})\simeq 0.034$~fm, and
$\Delta R_{n}({}^{176}{\rm Yb})\simeq 0.052$~fm, respectively.

\section{Conclusions}
\label{sec:conclusions}
The neutron radius of ${}^{208}{\rm Pb}$, an observable highly 
sensitive to the density dependence of the symmetry energy, is
correlated to ground-state observables of neutron-rich nuclei.
We find that models with small neutron skins in $^{208}$Pb 
predict weak binding energies for the valence neutron orbitals 
in neutron-rich nuclei. Thus, models with the softest symmetry 
energy are the first ones to drip neutrons. Further, a tight 
correlation was found between the neutron skin of 
${}^{208}{\rm Pb}$ and the neutron radius of a variety of
elements (barium, dysprosium, and ytterbium) of possible
relevance to atomic parity violation.

The softening of the symmetry energy, which generates a thinner
neutron skin in ${}^{208}{\rm Pb}$, is accomplished by introducing
additional (isoscalar-isovector) terms into the effective Lagrangian.
The new terms represent an important addition to the relativistic
mean-field models, as they allow modifications to the neutron skin of
${}^{208}{\rm Pb}$ without compromising their success in reproducing a
variety of ground-state observables.  At first, it might seem
surprising that the neutron radius of ${}^{208}{\rm Pb}$, one of the
most studied nuclei both theoretically and experimentally, should be
so poorly known. Indeed, theoretical estimates place an uncertainty in
the neutron radius of ${}^{208}{\rm Pb}$ at about $0.3$~fm. This is in
contrast to its charge radius which is known---experimentally---to
exquisite accuracy. The reason for this mismatch is twofold. First,
electron scattering experiments, arguably the cleanest probe of
nuclear structure, are only sensitive to the proton distribution.
Second, best-fit models constrained to reproduce a variety of
nuclear observables, such as charge densities, binding energies, and
single-particle spectra, still predict a wide range of neutron radii
for ${}^{208}$Pb.

Fortunately, the measurement of the neutron radius of ${}^{208}$Pb
seems within reach. Indeed, the Parity Radius Experiment at the 
Jefferson Laboratory aims to measure the neutron radius in $^{208}$Pb 
accurately (at the 1\% level) and model independently via
parity-violating electron scattering. Such a measurement seems vital,
as knowledge of a single isovector observable is sufficient to place 
stringent constraints on the model dependence of symmetry energy.
Further, such a measurement will shed light on a variety of rich nuclear 
phenomena, ranging from the structure of neutron-rich nuclei to the 
structure of neutron stars.
%%%%%%%%%%%%%%%%%%%%%%%%%%%%%%%%%%%%%%%%%%%%%%%%%

\smallskip
This work was supported in part by DOE grant DE-FG05-92ER40750.

%%%%%%%%%%%%%%%%%%%%%%%%%%%%%%%%%%%%%%%%%%%%%%%%%%%%%%%%%%%%%%%%%

\begin{thebibliography}{99}
\bibitem{Ho01a} C.J. Horowitz and J. Piekarewicz,
                Phys. Rev. Lett.~{\bf 86}, 5647 (2001).
\bibitem{Ho01b} C.J. Horowitz and J. Piekarewicz,
                Phys. Rev. C.~{\bf 64}, 062802 (2001).
\bibitem{Ho02}  C.J. Horowitz and J. Piekarewicz,
                Phys. Rev. C.~{\bf 66}, 055803 (2002).
\bibitem{Ca02}  J. Carriere, C.J. Horowitz and J. Piekarewicz,
                {\tt nucl-th/0211015}. 
\bibitem{Ho01c} C. J. Horowitz, S. J. Pollock, P. A. Souder 
	        and R. Michaels, 
	        Phys. Rev. C~{\bf 63}, 025501 (2001). 
\bibitem{Mi99}  Jefferson Laboratory Experiment E-00-003, 
                Spokespersons R. Michaels, P. A. Souder 
	        and G. M. Urciuoli.
\bibitem{Sl02}  P. Slane, D.J. Helfund, and S.S. Murray,
	        Astrophys. J {\bf 571}, L45 (2002).
\bibitem{Do94}  J. Dobaczewski, I. Hamamoto, W. Nazarewicz, 
                and J.A. Sheikh,  
		Phys. Rev. Lett.~{\bf 72}, 981 (1994).
\bibitem{Ot01}  T. Otsuka, R. Fujimoto, Y. Utsuno, B. Alex Brown, 
                Michio Honma and T. Mizusaki, 
                Phys. Rev. Lett. {\bf 87}, 082502 (2001).
\bibitem{Sa99}  H. Sakurai {\it et al.,}
                Phys. Lett. B~{\bf 448}, 180.
\bibitem{Pi00}  J. Piekarewicz,
	        Phys. Rev. C~{\bf 62}, 051304(R) (2000);
                Phys. Rev. C~{\bf 64}, 024307 (2001).
\bibitem{Mu96}  H. M\"uller and B.D. Serot, 
	        Nucl. Phys. {\bf A606}, 508 (1996).
\bibitem{Se86}  B.D. Serot and J.D. Walecka, Adv. in Nucl. 
                Phys. {\bf 16}, J.W. Negele and E. Vogt, eds. 
                (Plenum, N.Y. 1986); Int. Jour. Mod. Phys. 
                {\bf E6}, 515 (1997).
\bibitem{La97}  G. A. Lalazissis, J. K\"onig and P. Ring, 
	        Phys. Rev. C~{\bf 55}, 540 (1997).
\bibitem{La99}  G. A. Lalazissis, S. Raman, and P. Ring,
		Atomic Data and Nuclear Data Tables {\bf 71},
		1 (1999).
\bibitem{Da03}  Pawel Danielewicz, {\tt nucl-th/0301050}.
\bibitem{Au95}  G. Audi and A.H. Wapstra,
	        Nucl. Phys. {\bf A595}, 409 (1995).
\bibitem{Vr87}  H. de Vries, C.W. de Jager, and C. De Vries,
	        Atomic Data and Nucl. Data Tables,
	        {\bf 36}, 495 (1987).
\bibitem{Fr95}  G. Fricke, C. Bernhardt, K. Heilig, L.A. Schaller,
                L. Schellenberg, E.B. Shera, and C.W. Dejager,
		Atomic Data and Nuclear Data Tables {\bf 60},
		177 (1995).
\bibitem{Wa90}  E.K. Warburton, J.A. Becker, and B.A. Brown,
	        Phys. Rev. C~{\bf 41}, 1147 (1990).
\bibitem{Re95}  Zhongzhou Ren, W. Mittig, Baoqiu Chen, and
                Zhongyu Ma,
	        Phys. Rev. C~{\bf 52}, R20 (1995).
\bibitem{Kr97}  A.T. Kruppa, P.-H. Heenen, H. Flocard, and
                R.J. Liotta
	        Phys. Rev. Lett~{\bf 79}, 2217 (1997).
\bibitem{La98}  G. A. Lalazissis, D. Vretanar, W. P\"oschl,
                and P. Ring
	        Nucl. Phys. {\bf A632}, 363 (1998).
\bibitem{Xi90}  Xiong Xiaxing, He Mouqi, Zhao Youyuan, and
                Zhang Zhiming,
		J. Phys. B~{23}, 4239 (1990);
		J. Phys. B~{23}, 4239 (1991).
\bibitem{De95}  David DeMille,
	        Phys. Rev. Lett~{\bf 74}, 4165 (1995).
\bibitem{Bu99}  D. Budker, in {\it Proceedings of WEIN-98},
                edited by P. Herczeg, C. M. Hoffman, and
		H.V. Klapdor-Kleingrothaus (World Scientific, 
		Singapore, 1999).
\bibitem{Po92}  S.J. Pollock, E.N. Fortson, and L. Wilets
	        Phys. Rev. C~{\bf 46}, 2587 (1992).
\bibitem{Ch93}  B.Q. Chen and P. Vogel,
	        Phys. Rev. C~{\bf 48}, 1392 (1993).
\bibitem{Mu99}  M.J. Ramsey-Musolf,
	        Phys. Rev. C~{\bf 60}, 015501 (1999).
\bibitem{Sc02}  S. Schramm, {\tt nucl-th/0210053}.
\bibitem{Fu02}  R.J. Furnstahl, 
                Nucl. Phys. {\bf A706}, 85 (2002).
\bibitem{Pi02}  J. Piekarewicz, 
                Phys. Rev. C~{\bf 66}, 034305 (2002).
\end{thebibliography}
\end{document}